\documentclass{PoS}
\usepackage[latin9]{inputenc}
\usepackage{amsmath}
\usepackage{amssymb}
\usepackage{textcomp}
\usepackage{graphicx}
\DeclareGraphicsExtensions{.pdf,.eps}
\usepackage{subfig}
\usepackage{url}
\usepackage{xcolor}

\title{PDFSense: Mapping the PDF sensitivity of future facilities (HL-LHC, LHeC, and EIC)}

\ShortTitle{PDFSense: PDFs at future facilities}

\author{T.~J.~Hobbs$^{1,2,3}$,\, Bo-Ting~Wang$^1$,\, \speaker{Pavel~M.~Nadolsky}$^1$,\, and\, Fredrick~I.~Olness$^1$\\
        $^1$Department of Physics, Southern Methodist University, Dallas, TX 75275-0175, USA\\
        $^2$EIC Center at Jefferson Lab, 12000 Jefferson Ave, Newport News, VA 23606, USA\\
        $^3$E-mail: \email{tjhobbs@smu.edu}}

\abstract{
	Particle and nuclear physics are moving toward a new generation of
	experiments to stress-test the Standard Model (SM), search for novel degrees of freedom,
	and comprehensively map the internal structure of hadrons.
	Due to the complex nature of QCD and wide array of past, present, and possible future
	experiments, measurements taken at these next-generation facilities will inhabit an expansive
	space of high-energy data. Maximizing the impact of each future collider program
	will depend on identifying its place within this sprawling landscape.
	As an initial exploration, we use the recently-developed \texttt{PDFSense} framework
	to assess the PDF sensitivity of two future high-energy facilities --- the high-luminosity
	upgrade to the LHC (HL-LHC) and the Large Hadron-electron Collider (LHeC) proposal --- as well as the
	electron-ion collider (EIC) proposed to map the few-GeV quark-hadron transition region. We report that
	each of these experimental facilities occupies a unique place in the kinematical parameter space
	with specialized pulls on particular collinear quantities. As such, there is a clear complementarity
	among these programs, with an opportunity for each to mutually reinforce and inform the others.
	}

\dedicated{SMU-HEP-19-11}

\FullConference{XXVII International Workshop on Deep-Inelastic Scattering and Related Subjects - DIS2019\\
		8-12 April, 2019\\
		Torino, Italy}
\begin{document}
\section{Introduction}
\label{sec:intro}
\hyphenpenalty=1000
Particle and nuclear physics today find themselves at an important crossroads. At high
energies, the LHC has made tremendous progress in completing the Standard Model (SM) with the
recent discovery of the Higgs boson and ongoing tests of the SM. Meanwhile, at more intermediate
energies, experimental programs at JLab, RHIC, and a number of other facilities have
made strides in refining our understanding of hadronic bound states and of the
properties of nuclei.
Despite these advances, numerous questions about fundamental physics remain.
Among these are the quest to unravel the interactions of the Higgs with other SM particles,
the origin of the matter-antimatter asymmetry of the universe,
and the the fact that the exact nature of dark matter remains unidentified.
Meanwhile, our ability to systematically relate the bulk properties of strongly-interacting matter with the quark-gluon
dynamics of the underlying theory (QCD) remains limited. For these reasons, a next generation
of experiments has been proposed with the objective of making decisive advances on these
complementary fronts.

On the HEP phenomenology side, the planned High-Luminosity LHC (HL-LHC) \cite{Apollinari:2017cqg} and a possible
high-energy DIS collider, the Large Hadron-electron Collider (LHeC) \cite{AbelleiraFernandez:2012cc} are expected to clarify
our understanding of the electroweak sector, high-energy QCD, and perform more sensitive
collider searches for new physics. In particular, the HL-LHC is expected to achieve
percent-level precision in measurements of the Higgs couplings, as well as very precise
determinations of electroweak observables like the $W$-boson mass, $M_W$,
and weak-mixing angle, $\sin^2\theta_W$. At the same time, at medium energies near the
quark-hadron transition region at $Q\! \sim\! \mathrm{few\,\, GeV}$, an Electron-Ion Collider
(EIC) \cite{Accardi:2012qut} has been proposed for tomographic exploration of 
QCD bound states by essentially probing the multi-dimensional
{\it wave function} of the nucleon and other hadrons or nuclei, a goal which necessitates
the collection of enormous amount of data.
Progress along the ``energy frontier'' and in the quark-hadron transition region therefore entails the
accumulation of very large, $\mathcal{O}(1\, \mathrm{ab}^{-1})$,
data sets. This enormous quantity of data will have important impact
on phenomenology at various scales.  For example, high-precision measurements
at an EIC will have important phenomenological implications for high-energy 
measurements, while explorations at very high scales carried out at HL-LHC or LHeC
can provide constraints to nucleon structure observables.

Due to the possible syntheses and complementary impacts among these different programs spanning the
energy spectrum, detailed accounting of possible overlaps and different pulls originating with
each proposal is an urgent necessity.
We explore this issue for unpolarized proton PDFs --- the essential nonperturbative input into
theoretical predictions of HEP observables at the LHC.
In particular, we concentrate on the pulls of pseudodata representative of each of the
programs noted above. To perform this analysis, we use the \texttt{PDFSense} framework
developed in Ref.~\cite{Wang:2018heo}, following the conventions established therein.
Additional details, including possible synergies between precise EIC
measurements and lattice QCD calculations --- discussed in passing here ---
can be found in Ref.~\cite{Hobbs:2019gob}. These proceedings may be read in
parallel with the talk upon which they are based, given as Ref.~\cite{proconline}.
%
\section{Future HEP programs and the nucleon's collinear structure}
\label{sec:HEP}
As of the time of this writing, the HL-LHC represents a planned upgrade
to the LHC, with the ultimate goal being a factor $\sim\!5$ improvement upon the
LHC's instantaneous luminosity, generating an integrated luminosity as large as 
$\mathcal{L} = 3\, \mathrm{ab}^{-1}$. Like the LHC Run 2, the HL-LHC will measure
$pp$ events at $\sqrt{s} = 14$ TeV, probing a wide range of parton momentum
fractions, \linebreak $5 \cdot 10^{-5} < x < 1$, at large factorization scales $\mu$.
The LHeC proposal, on the other hand, envisions the incorporation of a charged lepton
beam to scatter $E_{e^{\pm}} = 60$ GeV $e^\pm$ off protons in the LHC main hadron ring. This
scenario would result in a high-luminosity collider capable of reaching
unprecedented (for DIS) energies of $\sqrt{s} = \sqrt{4 E_{e^\pm} E_p}  = 1.3$ TeV;
in turn, this would afford access to very low $x \gtrsim 5 \cdot 10^{-6}$.

\begin{figure}
\includegraphics[clip,width=1.05\textwidth]{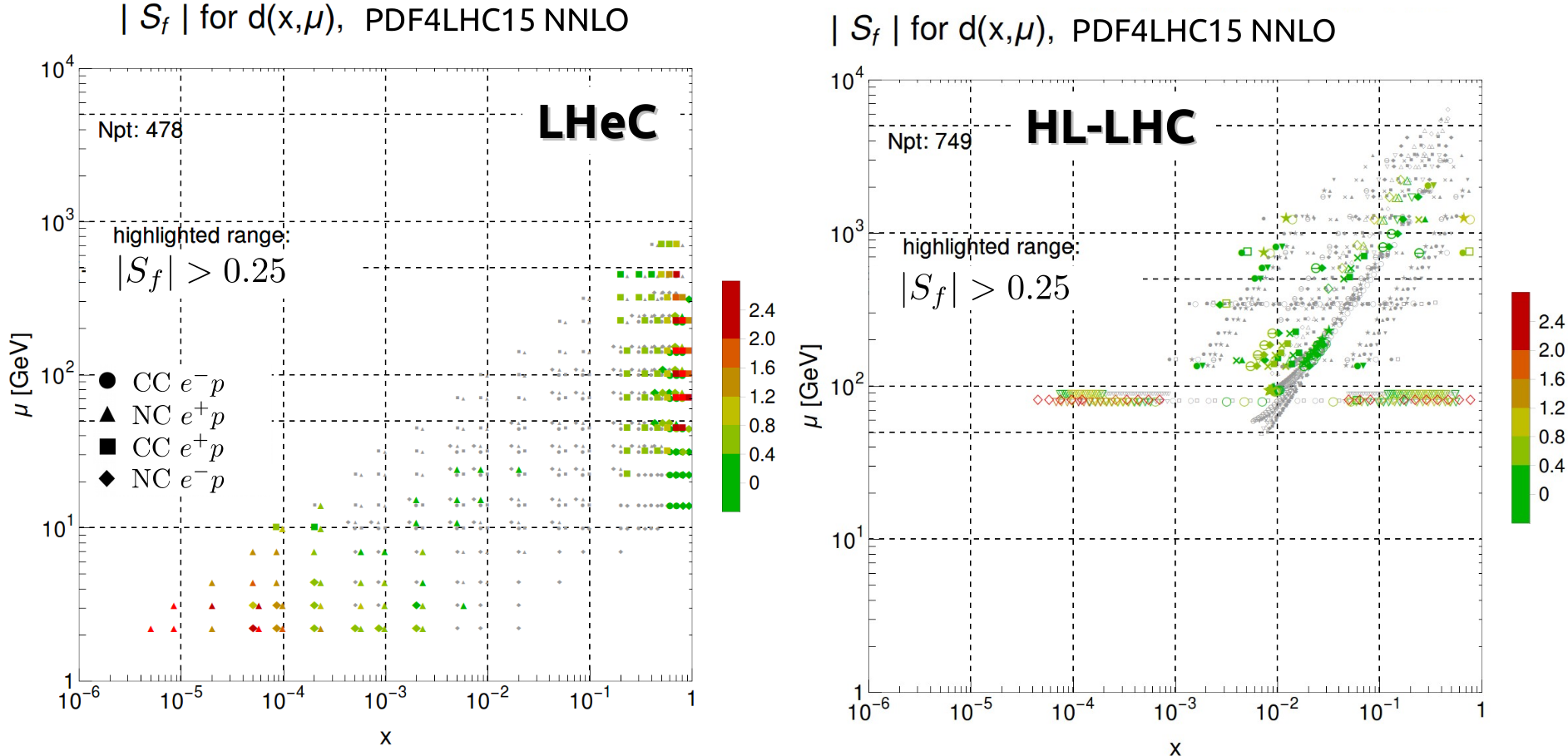}
	\caption{
	Future HEP experiments like the HL-LHC and possible LHeC proposal can have substantial PDF
	sensitivity, as shown here for the PDF4LHC15 $d(x,\mu)$ distribution computed according to the
	conventions in Ref.~\cite{Wang:2018heo}. The panels display the $x$- and $\mu$-dependent
	sensitivity of pseudodata for LHeC (left panel) and HL-LHC (right panel).
	}
\label{fig:HEP}
\end{figure}

The HL-LHC and LHeC programs are differentiated by the physics they would be adapted to explore.
We elucidate their differences by
implementing an assumed $\mathcal{L}=100\,\mathrm{fb}^{-1}$ of unpolarized electron/positron
scattering pseudodata from Ref.~\cite{LHeCdata}, as well as NNLO theory predictions
for HL-LHC pseudodata \cite{Khalek:2018mdn} at $\mathcal{L} = 3\, \mathrm{ab}^{-1}$
collected over a varied set of typical LHC processes and experiments; the resulting sensitivities
can then be directly compared in the \texttt{PDFSense} framework \cite{Wang:2018heo}.
The complementary, mostly non-overlapping regions of leading kinematic sensitivity for the HL-LHC vs.~LHeC can be
visualized by comparing kinematic distributions of experimental data points in the $(x,\mu)$ plane. Figure~\ref{fig:HEP} plots the sensitivity $|S_f|$ of the
LHeC (left panel) and HL-LHC (right panel) pseudodata to the $d$-quark distribution, $d(x,\mu)$, as
estimated using PDF4LHC15 NNLO PDFs \cite{Butterworth:2015oua}. 
We show this figure as a representative example of the differentiated pulls on the PDFs by the
LHeC and HL-LHC pseudodata. Similar comparisons can be made between HL-LHC and LHeC for PDFs of the gluon and other quark
flavors; a subset of these are shown in Slides 14-17 of Ref.~\cite{proconline}, and we refer the interested reader to those plots.

For the HL-LHC information in the right panel, the key strength is the coverage at the highest energy scales,
$\mu\! \sim\! 10$ TeV, for intermediate $x \gtrsim 0.01$. 
Meanwhile, the LHeC pseudodata shown on the left probe PDFs at $x\!\lesssim\! 10^{-5}$, beyond the
lower reach of HERA, primarily via neutral-current (NC) exchanges. 
In addition, at $x > 0.1$, the  charged-current (CC) DIS reduced cross section has the form
\begin{equation}
\sigma_{r,\,\mathrm{CC}}^{e^{+}p}  =\frac{Y_{+}}{2}W_{2}^{+}\mp\frac{Y_{-}}{2}xW_{3}^{+}-\frac{y^{2}}{2}W_{L}^{+}
                                 \simeq[1-y]^{2}\,x(d+s)+x(\overline{u}+\overline{c})\ ,
\label{eq:WCC}
\end{equation}
where $Y_{\pm}=1\pm(1-y)^{2}$, and from which one may infer $\sigma_{r,\,\mathrm{CC}}^{e^{+}p}\simeq xd\,[1-y]^{2}$
in the $x \to 1$ limit. The combination of NC and CC DIS at the LHeC would 
enjoy extensive coverage to perform the $u$ and $d$ separation in the high
$x > 0.1$ region over two decades of scales, $10^{1} < \mu < 10^{3}$ GeV. 

We thus see the complimentarity of the LHeC and the HL-LHC kinematic coverages for $d(x,\mu)$ in Fig.~\ref{fig:HEP}. It should be stressed that the sensitivity to the high-$x$ $d$-quark distribution at the LHeC comes without the potential ambiguities of a nuclear target (in contrast to typical extractions
of $F^n_2$ from DIS on the deuteron) and via a ``cleaner'' electroweak probe furnished by
lepton-nucleon DIS (rather than $pp$ collisions).  The HL-LHC pseudodata in the right panel would probe the intervening regions between the highest and lowest $x$ constrained by the LHeC, with the leading input especially coming from high-luminosity jet production at $\mu > 10^2$ GeV and $10^{-3} < x < 10^{-1}$. Precise data on $W^\pm$ hadroproduction extends the HL-LHC $d$-PDF sensitivity to high and low $x$, albeit over a comparatively narrower range of factorization scales. 

The NC LHeC pseudodata at $x<10^{-3}$ exhibits strong sensitivities to the gluon distribution (via Bjorken scaling violation) and the singlet quark PDF. 
Like the HERA program, the breadth of $x$ and $Q^2$ over which DIS observables would be accessible at the LHeC offers a strong
empirical ``lever-arm'' to test QCD evolution and the parameters of the theory in the perturbative region. 

\begin{figure}
\includegraphics[clip,width=0.5025\textwidth]{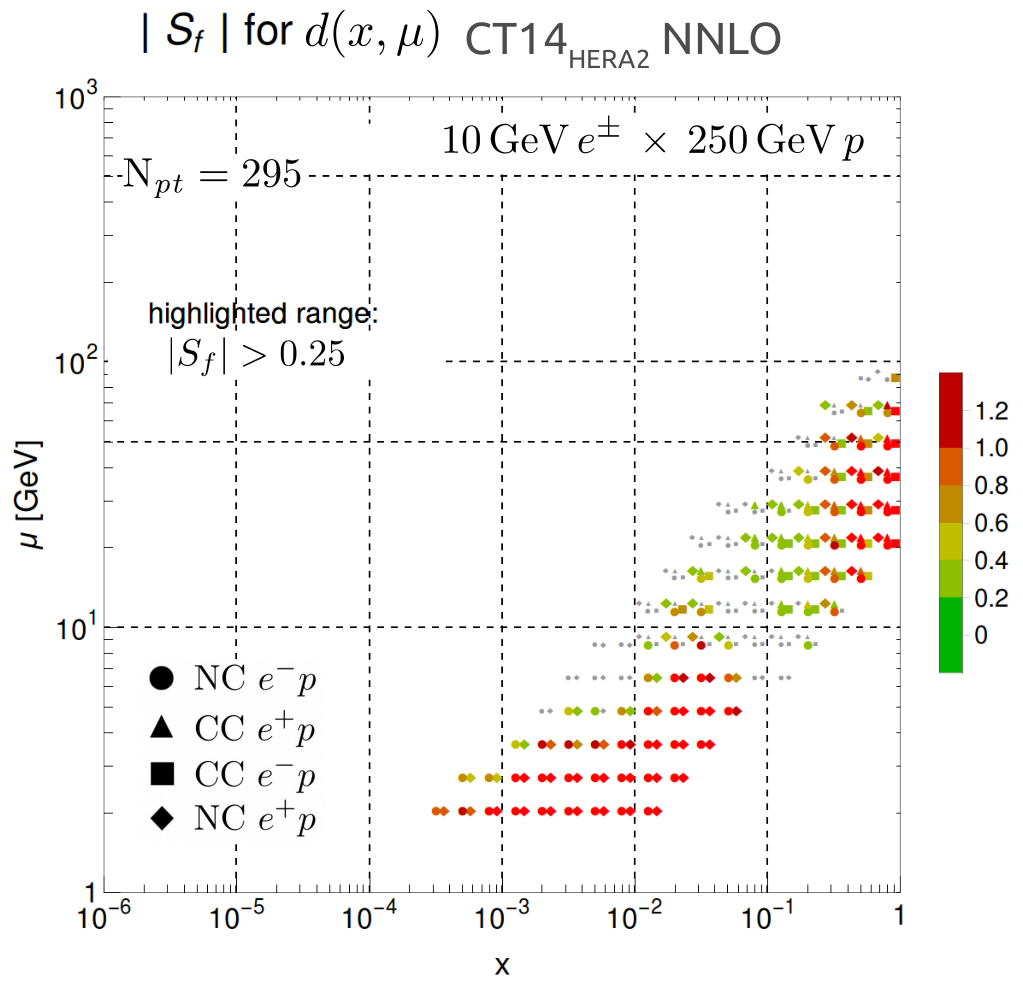}
\includegraphics[clip,width=0.5025\textwidth]{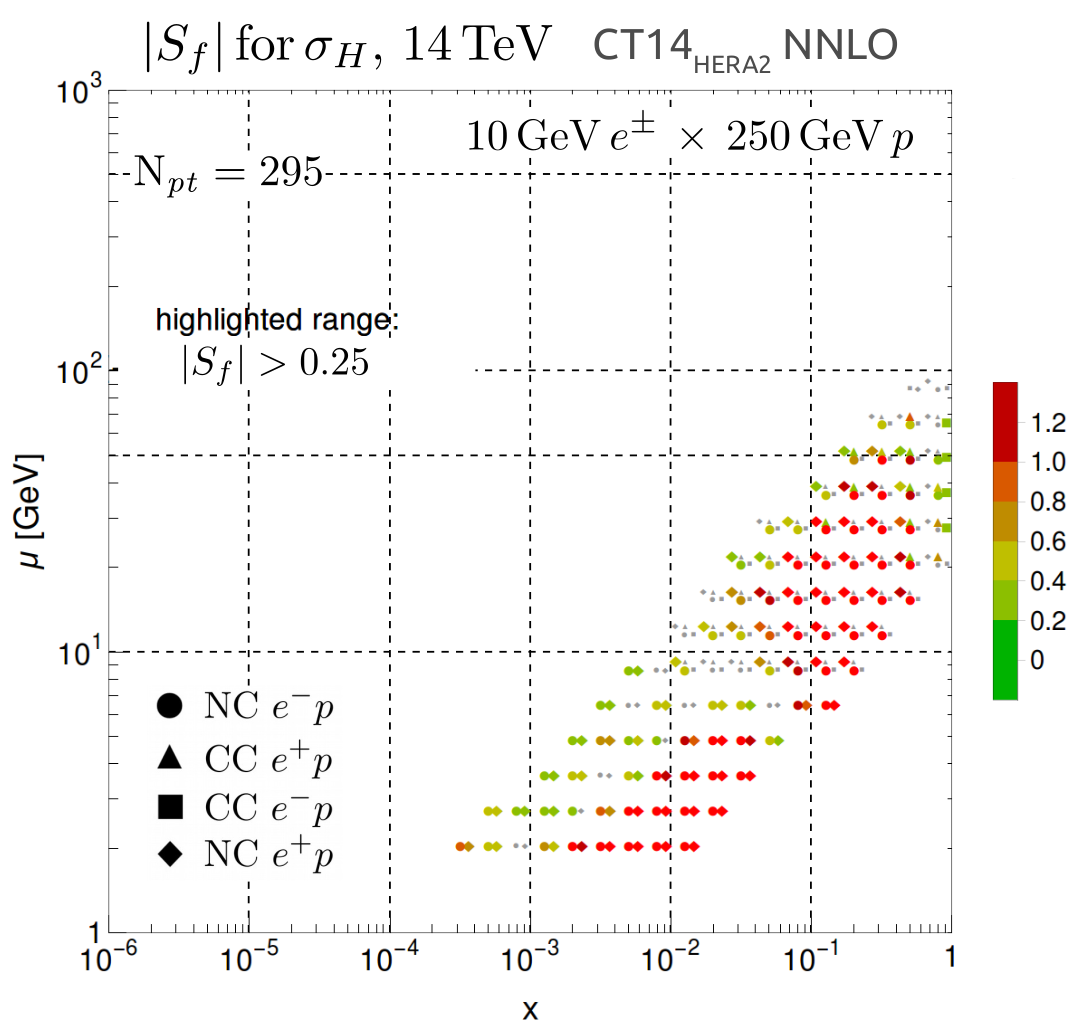}
	\caption{
		As a counterpart to the results plotted in Fig.~\ref{fig:HEP}, we show the
		EIC-like pseudodata sensitivity to $d(x,\mu)$ [left panel] as well as to
		$\sigma_H(\mathrm{14\,TeV})$, the total cross section for Higgs production at 14 TeV [right panel].
		The pseudodata assume an integrated luminosity of $\mathcal{L}=100\,\mathrm{fb}^{-1}$.
	}
\label{fig:EIC}
\end{figure}
It can also be instructive to compare the {\it aggregated} sensitivity, $\mathbf{S}_f \equiv \sum_i |S^i_f|$,
of each set of data to various PDFs to give an approximate sense for the total pull the pseudodata can be
expected to have in QCD analyses. By this figure of merit, the potential for mutual complementarity between
the flavor sensitivities of the HL-LHC and LHeC programs is strongly suggested \cite{proconline}. For the total
pull on the PDF4LHC15 gluon PDF, $\mathbf{S}_g$, we find
$\mathbf{S}^\mathrm{HL-LHC}_g =245 > \mathbf{S}^\mathrm{LHeC}_g = 151$, implying a potential
advantage for the HL-LHC in constraining the gluon. For the $d$-quark PDF plotted in Fig.~\ref{fig:HEP},
this balance is reversed, with a larger aggregated sensitivity for LHeC,
$\mathbf{S}^\mathrm{LHeC}_g =214 > \mathbf{S}^\mathrm{HL-LHC}_g = 171$. LHeC similarly
enjoys a modest advantage in unfolding nucleon strangeness, while there is close
parity in the sensitivity to the $\bar{d},\,\bar{u}$ distributions. It should be emphasized,
however, that the integrated luminosity assumed here for the LHeC pseudodata ($\mathcal{L}=100\,\mathrm{fb}^{-1}$)
is a fraction of what a full LHeC data-collection campaign could ultimately achieve. To the extent that
LHeC errors might be statistics-limited, these projections should be considered lower bounds.
%
%
%
%
\section{The HEP implication of an EIC}
\label{sec:EIC}
The scope of measurements to be undertaken at an EIC will have powerful implications for PDFs, and, by extension, it will 
impact upon HEP phenomenology at the LHC and elsewhere.
We highlight the impact on the unpolarized
nucleon PDFs that one might expect with a generic, $\sqrt{s} = 100$ GeV DIS
collider that is broadly consistent with the proposed design profiles for either the
JLab-based (JLEIC) or BNL-based (eRHIC) incarnations of the EIC.
In Fig.~\ref{fig:EIC}, we plot the sensitivity of EIC pseudodata
to the CT14HERA2 NNLO PDF \cite{Hou:2016nqm} for the $d$ quark in the left panel, as
well as the PDF-driven sensitivity to the high-energy Higgs production cross section,
$\sigma_H(14\,\mathrm{TeV})$, in the right panel. Here, 
an upper scale choice of $|S_f| = 1.2$ for highlighted points is chosen.
By the panels of Fig.~\ref{fig:EIC}, the powerful sensitivity of the EIC pseudodata 
generally surpasses that of the fixed-target
experiments that currently dominate the constraints on high-$x$ PDFs. The EIC will therefore
strongly constrain PDF dependence of HEP observables at moderate and large $x$,
including several in the Higgs and electroweak sectors like $M_W$ and $\sin^2\theta_W$.
As an emblematic example of this, the right panel of Fig.~\ref{fig:EIC}
shows the substantial EIC sensitivity to the total Higgs production cross section
at the LHC, $\sigma_H(14\,\mathrm{TeV})$. The constraints that a medium-energy machine like an
EIC would place on Higgs phenomenology stem from the predominance
of the $gg \to H$ fusion channel in $\sigma_H(14\,\mathrm{TeV})$.
In particular, $\sigma_H(14\,\mathrm{TeV}) = 62.1\, \mathrm{pb}$,
of which $88 \pm 4\%$ emanates from gluon fusion. While the leading
sensitivity to Higgs production at the LHC is expected to originate from the
``Higgs region'' at $\mu = m_H = 125$ GeV and $x \sim m_H/(14\, \mathrm{TeV}) \sim 0.01$,
QCD evolution connects the gluon PDF behavior at such $x$ and $\mu$ to the behavior
at the lower $\mu$ and higher $x$ that will be probed by the EIC.

Just as the EIC, {\it lattice QCD} calculations aim at
detailed understanding of the structure, spectrum, and interactions of the nucleon,
lighter hadrons, and their excitations. The recent analysis of 
Ref.~\cite{Hobbs:2019gob}, briefly summarized in Slides 18-25 of Ref.~\cite{proconline}, 
has explored the possibility of a future
synergy between phenomenology informed by the EIC data and lattice calculations
of integrated PDF moments and quasi-PDFs, $\widetilde{q}(x,\mu,P_z)$.
Cooperation between lattice studies and an EIC tomography
program would constrain nucleon PDFs necessary for HEP
phenomenology. Advances in lattice QCD techniques driven by
EIC-improved benchmarks could similarly feed-forward into lattice
calculations of, {\it e.g.}, branching ratios or quantities sensitive
to CP-violation.

{\it Conclusions}.
We reiterate our principle finding: the HL-LHC, EIC, and LHeC
each have unique and complementary access to a broad range of physics
as embodied by their PDF pulls. The HL-LHC will be distinguished by
its reach to the highest $\mu$ scales in a variety of $pp$ processes
with strong sensitivity to the gluon and $\bar{u},\,\bar{d}$
distributions. On the other hand, the LHeC would leverage its combination of
LHC energies with the DIS collider process to access very low 
and high $x$, with especially strong impact on the $g$, $u$, $d$,
and $s$ parton densities. The combination of the HL-LHC and LHeC constraints on PDFs across 
a wide range of $x$ and $\mu$ will be vital for the high-precision energy-frontier physics 
program that the HL-LHC will pursue. 
The complementarity we find between the HL-LHC and LHeC
is broadly consistent with the results recently reported in
Ref.~\cite{AbdulKhalek:2019mps}.
Lastly, by pursuing hadron tomography with extremely high precision using
polarized beams, an EIC would supersede the
bulk of fixed-target DIS experiments, providing critical information
needed to disentangle the nucleon's nonperturbative structure.
%
%
%
\subsection*{Acknowledgments}
We are grateful to Yulia Furletova and Jun Gao
for providing pseudodata for EIC and HL-LHC, respectively. We also
thank Alberto Accardi, Rik Yoshida, and CTEQ members for helpful discussions.
This work was supported by the U.S. Department of Energy under Grant No. DE-SC0010129.
The research of TJH is supported by an EIC Center@JLab Fellowship.


\providecommand{\href}[2]{#2}\begingroup\raggedright\endgroup


\end{document}